\documentclass[a4paper,11pt]{article}
\usepackage{pos}

\title{Status and prospects of Muon g-2 experiment}

\author*[a, b, 1]{Marin Karuza}

\affiliation[a]{INFN Sezione di Trieste,\\
  Via Valerio 2, Trieste, Italy}

\affiliation[b]{Faculty of Physics, University of Rijeka,\\
Radmile Matejcic 2, Rijeka, Croatia}
\note{For the Muon g-2 collaboration.}

\emailAdd{mkaruza@uniri.hr}

\abstract{This article reviews the muon g-2 experiment, a cornerstone in precision tests of the Standard Model of particle physics. The experiment measures the anomalous magnetic moment of the muon with unprecedented accuracy, seeking potential discrepancies between theoretical predictions and experimental results that might indicate physics beyond the Standard Model. We trace the evolution of this measurement from its beginnings at CERN in the 1960s to the current state-of-the-art experiment at Fermilab, highlighting the remarkable engineering achievements required to achieve parts-per-billion precision. Recent results from Runs 1-3 have achieved a systematic uncertainty of 70 ppb, exceeding design goals, while ongoing theoretical calculations continue to refine predictions. Despite these advances, the analysis remains statistics-limited, with continued data collection and novel experimental approaches at MUonE and J-PARC promising further insights into this fundamental physical quantity.}

\FullConference{Third Italian Workshop on the Physics at High Intensity (WIFAI2024)\\
 12-15 November 2024\\
 Bologna, Italy\\}

\begin{document}
\maketitle

\section{Introduction}
The muon, discovered in 1936 by Carl Anderson and Seth Neddermeyer \cite{AndersonNeddermeyer1937} while observing cosmic rays atop Pike's Peak, has become a crucial particle in our understanding of fundamental physics. This long-lived particle, often referred to as having a "Goldilocks mass," has proven to be an invaluable tool in probing the properties of the quantum vacuum and testing the Standard Model of particle physics. The term "Goldilocks mass" aptly describes the muon's unique position on the particle mass scale - at 207 times the mass of an electron yet only about 1/17 the mass of a tau lepton, it occupies a privileged middle ground that makes it especially valuable for precision experiments. Its mass is substantial enough to be significantly more sensitive to potential new physics effects than electrons, as heavier particles more readily interact with high-energy quantum fluctuations. Yet it remains light enough to possess a relatively long lifetime of 2.2 microseconds, giving scientists ample time to manipulate, store, and measure its properties before decay. Furthermore, the muon predominantly decays through a clean leptonic channel ($\mu^- \to e^- + \bar\nu_e + \nu_\mu$), with other decay modes suppressed by at least five orders of magnitude, creating an unmistakable experimental signature that provides excellent information about the muon's spin orientation at the moment of decay. This combination of mass, lifetime, and decay characteristics makes the muon neither too heavy nor too light, neither too short-lived nor too stable, but "just right" for probing the deepest questions in fundamental physics.
At the heart of the g-2 experiment lies the concept of the magnetic dipole moment, a property shared by particles like electrons and muons which makes them similar to tiny bar magnets. In quantum mechanics, the Dirac equation \cite{Dirac1928} predicted that the gyromagnetic ratio (g) for these particles should be exactly 2. However, in 1948, Foley and Kusch measured the electron's g-factor and found it slightly larger than 2, with a value of 2.00238 \cite{KuschFoley1948}. This tiny deviation from 2 became the focus of incredibly precise measurements and theoretical calculations.
The modern g-2 experiment measures this deviation with unprecedented precision by observing muon spin precession in a magnetic field. The experiment utilizes a specific "magic" momentum where certain experimental effects conveniently cancel out, allowing for cleaner measurements of the fundamental physics involved.
The journey of g-2 measurements spans several decades and facilities, representing a steady progression in experimental precision and theoretical understanding. The initial experiments at CERN in the 1960s employed a novel technique of storing muons in a magnetic ring and observing their spin precession \cite{Charpak1962}. These pioneering measurements \cite{Bailey1968} achieved a precision of about 1\% and provided crucial early confirmation of Quantum Electrodynamics (QED), validating the theoretical framework that describes how light and matter interact. The CERN experiments  established the methodology that would be refined in subsequent decades, demonstrating that the measured anomalous magnetic moment agreed with theoretical calculations that included quantum loop corrections.
By the 1970s, improvements at CERN pushed the precision to parts per thousand \cite{Farley1966}, continuing to validate increasingly sophisticated QED calculations. The experimental approach evolved significantly when the experiment moved to Brookhaven National Laboratory (BNL) in the 1990s. The E821 experiment at BNL represented a quantum leap in precision \cite{Bennett2006}, employing superconducting magnets to create a more uniform field and advanced detector systems to track muon decay with greater accuracy. These innovations enabled measurements with precision reaching 1 part per million (ppm). Intriguingly, the BNL results suggested a potential discrepancy with Standard Model predictions, differing by more than two standard deviations. This tantalizing hint of possible new physics beyond the Standard Model provided strong motivation for the next generation experiment at Fermilab, with its goal of reducing uncertainty by a factor of four to definitively confirm or refute the apparent discrepancy \cite{Proposal2009}.
\section{Experiment}
\begin{figure}[h!]
\centering
\includegraphics[width=7cm]{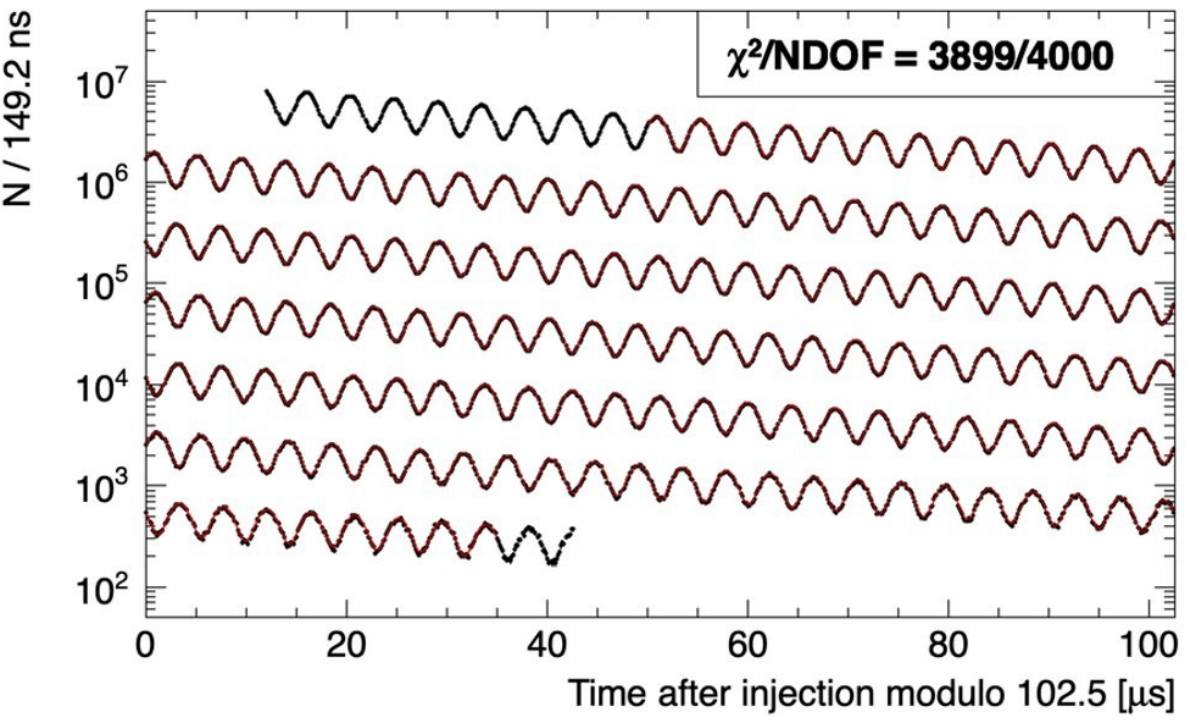}
\caption{Wiggle plot \cite{Albahri2021a}. An exponential decay modulated by the muon spin precession.}
\label{wiggle}
\end{figure}

The idea behind the experiment is in principle very simple. Polarized muons are stored in a magnetic storage ring, where they circulate while their spins precess. The principal measurement comes from detecting decay positrons (for positive $\mu$
) or electrons (for negative $\mu$) as they emerge from muon decays around the ring. The number of particles detected as a function of time gives a characteristic 'wiggle plot' shown in Fig. \ref{wiggle}, where the frequency of these oscillations provides the information about the magnetic anomaly,
\begin{equation}
\vec{\omega_a} = \frac{q}{m_\mu}\left[a_\mu\vec{B} - a_\mu\left(\frac{\gamma}{\gamma+1}\right)(\vec{\beta}\cdot\vec{B})\vec{\beta} - \left(a_\mu - \frac{1}{\gamma^2-1}\right)\vec{\beta}\times\frac{\vec{E}}{c}\right]
\end{equation}
This key frequency is basically the difference between the cyclotron frequency, $\omega_c = \frac{qB}{m_\mu}$ and the spin precession frequency, $\omega_s = \frac{g_\mu q}{2m_\mu}B$, when measured at the so-called "magic momentum" where the contribution of the third term becomes negligible since the coefficient in front of the electric field becomes zero and the  $\vec B$ is assumed orthogonal to $\beta$. At this specific momentum, approximately 3.09 GeV/\textit{c} and with above assumptions the difference gives the anomalous precession frequency, $\omega_a = \omega_s - \omega_c = \frac{a_\mu q}{m_\mu}B$, where $a_\mu = \frac{g_\mu-2}{2}$ is the anomalous magnetic moment. It's important to note that the spin precession frequency includes contributions from both the intrinsic magnetic moment interaction with the field and the Thomas precession,  $\omega_{Thomas} = -\frac{q}{m_\mu}\frac{\gamma-1}{\gamma}B$.

The "wiggle plot" shows number of detected decay positrons/electrons, above energy threshold, as a function of time, with an oscillation at frequency $\omega_a$ modulated by the exponential decay of the muon population,
\begin{equation}
N(t) = N_0 e^{-t/\tau} [1 + A \cos(\omega_a t + \phi)]
\end{equation}
where $\tau$ is the dilated muon lifetime, $A$ is the asymmetry parameter, and $\phi$ is a phase constant.

\section{Status}
The Fermilab Muon g-2 collaboration has published  the combined results from runs 1-3 \cite{Albahri2023, Abi2021, Aguillard2024}, significantly strengthening the case for a potential discovery of physics beyond the Standard Model. These measurements of the muon's anomalous magnetic moment now show a discrepancy exceeding 5 sigma when compared to the theoretical predictions in the "White Paper" consensus value \cite{Aoyama2020}. 
The new data not only confirms Run 1 results and the earlier Brookhaven National Laboratory (BNL) results from the early 2000s but does so with substantially improved precision and under different experimental conditions, addressing concerns that the original discrepancy might have been due to systematic effects specific to the BNL setup.
A cornerstone of this experimental achievement has been the meticulous study and accounting of systematic effects. The measured value requires a complex system of corrections represented by the formula:
\begin{equation}
\frac{\omega_a}{\omega_p} = \frac{\omega_a^m}{\omega_p^m} \times \frac{1+C_e+C_p+C_{pa}+C_{dd}+C_{ml}}{1+B_k+B_q}
\end{equation}

The total correction amounts to 622 parts per billion (ppb), primarily dominated by the electric field and pitch effects. These corrections can be categorized as follows:
\begin{itemize}
    \item Kinematic Corrections
\begin{itemize}
    \item $C_e$: Accounts for the second-order effects of the electric field on spin precession. The presence of electric fields in the storage ring, necessary for vertical focusing of the muon beam, introduces a relativistic correction to the spin precession frequency;
\item $C_p$: Addresses the "pitch" correction due to the vertical betatron oscillations (up and down motion) of muons in the storage ring. These oscillations cause the muon spin to precess slightly slower than predicted by the simplified theoretical equation.
\end{itemize}
    \item Detector and Beam Dynamics Corrections
\begin{itemize}
    \item $C_{pa}$: The phase-acceptance correction accounts for the correlation between a muon's initial phase and its probability of being detected;
    \item $C_{dd}$: The differential decay correction handles the time-dependent changes in the detected decay positron energy spectrum;
    \item $C_{ml}$: Corrects for muon losses during the measurement period, which could otherwise bias the precession frequency measurement.
\end{itemize}
\item Magnetic Field Perturbations
\begin{itemize}
    \item $B_k$: Corrects for transient magnetic fields created by eddy currents in the kicker magnets used to inject muons into the storage ring;
    \item $B_q$: Accounts for magnetic field perturbations from vibrations in the quadrupole focusing system.
\end{itemize}
\item Robust Analysis Methodology\\
To ensure the integrity of the results and eliminate potential experimenter bias, the collaboration implemented a comprehensive blind analysis strategy. This involved:
\begin{itemize}
\item Six independent analysis groups working in parallel;
\item Six distinct methodological approaches to extract the precession frequency;
\item Concealment of key calibration constants until all analyses were complete;
\item Rigorous cross-checking of all systematic uncertainty estimates.
\end{itemize}
This blind analysis framework represents the gold standard in experimental particle physics, providing strong assurance that the observed anomaly is not an artifact of the analysis procedure.
\end{itemize}
A particularly compelling aspect of these results is that they were obtained after significant experimental improvements between runs. The Fermilab experiment, while using the same storage ring as BNL, features:
\begin{itemize}
\item Enhanced magnetic field uniformity and measurement techniques;
\item Improved detector systems with higher granularity and better timing resolution;
\item More sophisticated beam delivery and muon injection systems;
\item Advanced calibration procedures and monitoring systems.
\end{itemize}
The consistency of results across these improved experimental configurations strongly suggests that the observed discrepancy with theory is a genuine physical effect rather than an experimental artifact.
The combined measurement from runs 1-3 yields a muon anomalous magnetic moment of $a_{\mu}(\text{FNAL}) = 116592059(22) \times 10^{-11}$, with a precision of 0.20 parts per million (ppm). This result is remarkably consistent with the previous BNL measurement of $a_{\mu}(\text{BNL}) = 116592089(63) \times 10^{-11}$, yielding a world-average experimental value of $a_{\mu}(\text{Exp}) = 116592061(18) \times 10^{-11}$. When compared to the 2020 White Paper prediction of $a_{\mu}(\text{SM}) = 116591811(43) \times 10^{-11}$, the difference $\Delta a_{\mu} = a_{\mu}(\text{Exp}) - a_{\mu}(\text{SM}) = 250(48) \times 10^{-11}$
 represents a statistically significant $5.2\sigma$ discrepancy. This level of disagreement between theory and experiment is striking, especially considering that the systematic uncertainty in the Fermilab measurement has been reduced to an unprecedented 70 ppb, substantially better than the original design goal. The consistency between two generations of experiments, along with the sharp reduction in both statistical and systematic uncertainties, lends considerable weight to the possibility that this discrepancy represents a genuine window into new physics beyond the Standard Model.

However, recent high-precision Lattice QCD calculations of the hadronic vacuum polarization contribution \cite{Hoyle2024, Borsanyi2021}  suggest a significantly higher value that would close much of the gap between theory and experiment. If these lattice calculations are confirmed to be correct, the enticing prospect of new physics explanations for the g-2 anomaly would diminish substantially since $\Delta a_\mu = 42(42) \times 10^{-11}$ is significantly reduced. Nevertheless, a profound mystery would remain as to why the data-driven dispersive approach, based on electron positron annihilation measurements  \cite{Davier2017, Keshavarzi2018} and considered the gold standard for decades, would have yielded consistently lower predictions. This tension between different theoretical approaches represents a fascinating puzzle in itself and may ultimately reveal important insights about QCD calculations or experimental techniques in precision physics.

\section{Prospects}
While data collection at Fermilab's Muon g-2 experiment has concluded, significant work remains in advancing our understanding of this critical measurement. The immediate task is finalizing the analysis of all collected data containing approximately 150 billion events. In particularly dataset from Runs 4, 5, and 6 is about three times larger than one from Runs 1 to 3. These results will be incorporated into comprehensive publications that present the full dataset findings, accompanied by auxiliary papers exploring specific aspects of the experiment in greater detail. Beyond the current experiment, the field is advancing along multiple fronts. The MUonE experiment \cite{Abbiendi2019} represents an innovative approach that will measure hadronic vacuum polarization (HVP) using the "space-like" method, which is based on the precise measurement of the Coulomb scattering angle of the muon on target electron. Unlike traditional, "time-like" measurement this technique provides a completely independent determination of the HVP, potentially offering insight into the discrepancy between theory and experiment. This provides an independent way to obtain this critical contribution to the muon g-2 value, potentially offering new insights into the persistent discrepancy between theory and experiment. At J-PARC  \cite{Abe2019}, scientists are taking a fundamentally different experimental approach by using low-energy muons without electric fields. This design choice eliminates certain systematic uncertainties present in previous experiments, creating yet another independent measurement path. These diverse experimental approaches reflect the physics community's commitment to resolving one of the most intriguing questions in particle physics: whether the tension between theoretical predictions and experimental measurements of the muon g-2 value represents evidence of physics beyond the Standard Model. Each new approach brings unique strengths and different systematic considerations, creating a robust framework for testing our understanding of fundamental physics.

\section{Acknowledgments}
The Muon Experiment was performed at the Fermi National Accelerator Laboratory, a U.S. Department of Energy, Office of Science, HEP User Facility. Fermilab is managed by Fermi Forward Discovery Group, LLC, acting under Contract No. 89243024CSC000002. Additional support for the experiment was provided by the Department of Energy offices of HEP, NP, and ASCR (USA), the National Science Foundation (USA), the Istituto Nazionale di Fisica Nucleare (Italy), the Science and Technology Facilities Council (UK), the Royal Society (UK), the National Natural Science Foundation of China (Grant No. 12211540001, 12075151), MSIP, NRF and IBS-R017-D1 (Republic of Korea), the German Research Foundation (DFG) through the Cluster of Excellence PRISMA+ (EXC 2118/1, Project ID 39083149), the European Union Horizon 2020 research and innovation programme under the Marie Sk\l{}odowska-Curie grant agreements No. 101006726, No. 734303, and European Union STRONG 2020 project under grant agreement No. 824093 and the Leverhulme Trust, LIP-2021-01.


\begin{thebibliography}{99}

\bibitem{AndersonNeddermeyer1937}
C.~D.~Anderson and S.~H.~Neddermeyer,
``Note on the Nature of Cosmic-Ray Particles,''
Phys. Rev. \textbf{51}, 884-886 (1937).

\bibitem{Dirac1928}
P.~A.~M.~Dirac,
``The Quantum Theory of the Electron,''
Proc. Roy. Soc. Lond. A \textbf{117}, 610-624 (1928).

\bibitem{KuschFoley1948}
P.~Kusch and H.~M.~Foley,
``The Magnetic Moment of the Electron,''
Phys. Rev. \textbf{74}, 250-263 (1948).

\bibitem{Charpak1962}
G.~Charpak et al.,
``Measurement of the anomalous magnetic moment of the muon,''
Phys. Rev. Lett. \textbf{6}, 128-132 (1961).

\bibitem{Bailey1968}
J.~M.~Bailey et al.,
``The anomalous magnetic moment of the muon,''
Phys. Lett. B \textbf{28}, 287-290 (1968).


\bibitem{Farley1966}
F.~J.~M.~Farley and E.~Picasso,
``The muon (g-2) experiments,''
Ann. Rev. Nucl. Part. Sci. \textbf{29}, 243-282 (1979).

\bibitem{Bennett2006}
G.~W.~Bennett et al. [Muon g-2 Collaboration],
``Final Report of the Muon E821 Anomalous Magnetic Moment Measurement at BNL,''
Phys. Rev. D \textbf{73}, 072003 (2006).

\bibitem{Proposal2009}
Carey, R.M. et al. "The New (g-2) Experiment: A proposal to measure the muon anomalous magnetic moment to ±0.14 ppm precision." Fermilab-Proposal-0989, 2009.

\bibitem{Albahri2021a}
T.~Albahri et al. [Muon g-2 Collaboration],
``Measurement of the anomalous precession frequency of the muon in the Fermilab Muon g-2 experiment,''
Phys. Rev. D \textbf{103}, 072002 (2021).

\bibitem{Abi2021} 
B.~Abi et al. [Muon g-2 Collaboration],
``Measurement of the Positive Muon Anomalous Magnetic Moment to 0.46 ppm,''
Phys. Rev. Lett. \textbf{126}, 141801 (2021).

\bibitem{Albahri2023}
T.~Albahri et al. [Muon g-2 Collaboration],
``Measurement of the positive muon anomalous magnetic moment to 0.20 ppm,''
Phys. Rev. Lett. \textbf{131}, 161802 (2023).

\bibitem{Aguillard2024}
D.~P.~Aguillard et al. [Muon g-2 Collaboration],
``Detailed report on the measurement of the positive muon anomalous magnetic moment to 0.20 ppm,''
Phys. Rev. D \textbf{110}, 032009 (2024)

\bibitem{Aoyama2020}
T.~Aoyama et al.,
``The anomalous magnetic moment of the muon in the Standard Model,''
Phys. Rept. \textbf{887}, 1-166 (2020).

\bibitem{Hoyle2024}
C.~D.~Hoyle et al.,
``High precision calculation of the hadronic vacuum polarisation contribution to the muon anomaly,''
arXiv:2407.10913 (2024).

\bibitem{Borsanyi2021}
S.~Borsanyi et al.,
``Leading hadronic contribution to the muon magnetic moment from lattice QCD,''
Nature \textbf{593}, 51-55 (2021).

\bibitem{Davier2017}
M.~Davier, A.~Hoecker, B.~Malaescu, and Z.~Zhang,
``Reevaluation of the hadronic vacuum polarisation contributions to the Standard Model predictions of the muon $g-2$ and $\alpha(m^2_Z)$ using newest hadronic cross-section data,''
Eur. Phys. J. C \textbf{77}, 827 (2017).

\bibitem{Keshavarzi2018}
A.~Keshavarzi, D.~Nomura, and T.~Teubner,
``Muon $g-2$ and $\alpha(m^2_Z)$: a new data-based analysis,''
Phys. Rev. D \textbf{97}, 114025 (2018).

\bibitem{Abbiendi2019}
G.~Abbiendi et al. [MUonE Collaboration],
``Letter of intent: the MUonE project,''
CERN-SPSC-2019-026 / SPSC-I-252 (2019).

\bibitem{Abe2019}
M.~Abe et al.,
``A new approach for measuring the muon anomalous magnetic moment and electric dipole moment,''
PTEP \textbf{2019}, 053C02 (2019).




























\end{thebibliography}
\end{document}